\documentclass[twoside]{article}

\usepackage{float}
\usepackage{color}
\usepackage{greekbf}
\usepackage{float}
\usepackage{amsmath,amssymb}
\usepackage{mathrsfs}
\usepackage{enumerate}
\usepackage{srcltx}
\usepackage{mathtools}
    \mathtoolsset{showonlyrefs}
\usepackage{graphicx, psfrag}
\usepackage{setspace}
\usepackage{hyperref}

\hypersetup{bookmarks=false,
colorlinks=false,
linkcolor=black,
urlcolor=black,
pagebackref=true,
pdfstartview={FitH},
bookmarksopenlevel=2
}

\DeclareMathAlphabet{\mathbf}{OT1}{cmr}{bx}{it}
\definecolor{red}{rgb}{0.9,0,0}
\definecolor{blue}{rgb}{0.2,0.2,0.8}
\definecolor{green}{rgb}{0.0,0.5,0.2}
\definecolor{darkblue}{rgb}{0.2,0.2,0.5}
\definecolor{orange}{rgb}{1,0.5,0}
\definecolor{pink}{rgb}{0.96,0.5,0.46}
\definecolor{lblue}{rgb}{0.18,0.74,1}
\definecolor{cyan}{rgb}{0,0.8,0.8}

\newcommand {\Ev} {\mathbf{E}^{\rm vk}}
\newcommand {\El} {\mathbf{E}}
\newcommand {\vb} {\mathbf{v}}
\newcommand {\wl} {{w}^\ell}
\newcommand {\vbl} {\mathbf{v}^\ell}
\newcommand {\eb}{\mathbf{e}}

\newcommand {\pb} {\mathbf{p}}
\newcommand {\qb} {\mathbf{q}}

\newcommand {\ub} {\mathbf{u}}
\newcommand {\xb} {\mathbf{x}}

\newcommand {\Ab} {\mathbf{A}}

\newcommand {\Pb} {\mathbf{P}}

\newcommand {\Cc}  {\mathcal{C}}
\newcommand {\Dc}  {\mathcal{D}}

\newcommand {\Uc}  {\mathcal{U}}

\newcommand {\Zc}  {\mathcal{Z}}

\newcommand {\db} {\mathbf{d}}

\renewcommand {\sb} {\mathbf{s}}

\DeclareMathOperator{\Thz}{\overset{(z)}{\Theta}}
\DeclareMathOperator{\Thc}{\overset{(c)}{\Theta}}
\DeclareMathOperator{\Psic}{\overset{(c)}{\Psi}}
\DeclareMathOperator{\Psiz}{\overset{(z)}{\Psi}}
\DeclareMathOperator{\varthetat}{\overset{(2)}{\vartheta}}
	\DeclareMathOperator{\psit}{\overset{(2)}{\psi}}

\DeclareMathOperator{\vfb}{\boldsymbol{\mathfrak{v}}^\ell}
\DeclareMathOperator{\wf}{\mathfrak{w}^\ell}

\DeclareMathOperator{\tr}{\textrm{tr}}

\newcommand{\Dl}[1]{D^\ell_{\pb_{#1}}}
\newcommand{\Dlo}[1]{D^\ell_{-\pb_{#1}}}

\begin{document}
	
\doublespacing

\title{\vspace{-3cm} {\bf  An atomistic-based \\F\"{o}ppl--von K\'{a}rm\'{a}n model for graphene }}

\author{
Cesare Davini$^1$\!\!\!\!\! \and  Antonino Favata$^2$\!\!\!\!\! \and Roberto Paroni$^{3}$
}

%
%\date{\today}

%%% ----------------------------------------------------------------------
\maketitle
%%% ----------------------------------------------------------------------

\vspace{-1cm}
\begin{center}
	{\small
		$^1$ Via Parenzo 17, 33100 Udine\\
		\href{mailto:cesare.davini@uniud.it}{cesare.davini@uniud.it}\\[8pt]
		$^2$ Department of Structural and Geotechnical Engineering\\
		Sapienza University of Rome, Rome, Italy\\
		\href{mailto:antonino.favata@uniroma1.it}{antonino.favata@uniroma1.it}\\[8pt]

		$^3$ Dipartimento di Ingegneria Civile e Industriale\\
	Universit\`{a} di Pisa, Pisa, Italy\\
		\href{mailto:roberto.paroni@unipi.it}{roberto.paroni@unipi.it}
	}
\end{center}

\pagestyle{myheadings}
\markboth{C.~Davini, A.~Favata, R.~Paroni }
{An atomistic-based F\"{o}ppl--von K\'{a}rm\'{a}n model for graphene}

\vspace{-0.5cm}
\section*{Abstract}

We  deduce a non-linear continuum model of graphene  for the case of finite out-of-plane displacements and  small in-plane deformations. On assuming that the lattice interactions are governed by the Brenner's REBO potential of 2nd generation and that self-stress is present, we introduce discrete strain measures accounting for up-to-the-third neighbor interactions. The continuum limit turns out to depend on an average (macroscopic) displacement field and  a relative {\it  shift displacement}  of the two Bravais lattices that give rise to the hexagonal periodicity. On minimizing the energy with respect to the shift variable, we formally determine a continuum model  of F\"{o}ppl--von K\'{a}rm\'{a}n type, whose constitutive coefficients are given in terms of the atomistic interactions.

\vspace{1cm}
\noindent {\bf Keywords}: Graphene, Continuum Modeling, F\"{o}ppl--von K\'{a}rm\'{a}n equations.

\tableofcontents

\section{Introduction}\label{sec:INTROD}

For its extraordinary mechanical, electrical, and thermal properties graphene is one of the most studied materials of the last two decades. Its discovery gave the occasion to renew a classical debate on the stability of 2D materials in nature \cite{landau1980, mermin1968}, and opened the way to the discussion of many theoretical issues. From the onset it was clear that the applications in various fields of technology might have been revolutionary. Though it's fair to say that graphene's potentialities are far from being fully explored and exploited, and remain the object of an intensive study, as much as it would be difficult to give an account of the huge literature on the subject. For a general picture of the subject and the new technological applications see the review by Ferrari et al. \cite{Ferrari2014}.

The availability of macroscopic models is crucial to design applications and   experiments. The simplest models of structural mechanics such as membranes, plates, and shells have been often adopted in the past; in some cases, they have been assumed as \textit{a priori} models and  the relevant constitutive constants have been estimated from {\it ab initio} or Molecular Dynamics simulations. Huang et al. \cite{huang2006} highlight how this has resorted to a bit of a stretch in some cases and led to paradoxical conclusions. More recent contributions start from atomistic analyses based on appropriate constitutive assumptions  on the interatomic potentials to obtain the continuum models of structural mechanics.

 Thus, Lu and Huang \cite{luHuang2009} estimated the elastic modulus and  bending stiffness of a graphene sheet from Molecular Mechanics calculations by  considering the one-dimensional stretching and the rolling on cylinders of various radii of a rectangular piece of graphene, and assuming that the interatomic forces are ruled by  Brenner's REBO potential of the second type. In this way they obtained values of the elastic constant  and bending stiffness that closely agree with those found by Kudin et al. \cite{kudin2001} from {\it ab initio} calculations. These values are twice as much as  those found by Arroyo and Belitschko \cite{arroyo2004} in a paper where the dihedral contribution in Brenner's potential is taken into account; Arroyo and Belitschko also gave an atomistic-based membrane model for single layer crystalline films,  \cite{arroyo2002}. Finally, Davini  \cite{davini2014} deduced a 2D continuum model for the in-plane deformations of a graphene sheet within the framework of $\Gamma$-convergence. 

The out-of-plane deformations have been considered by various authors.  In particular,  by exploiting a formal analysis  we  deduced a continuum model of a graphene sheet \cite{davini2017}, and provided explicit expressions for the {\it bending} and {\it Gaussian stiffness} by starting from the study of the lattice kinematics and assuming the reactive empirical bond-order potential (REBO) of 2nd generation by Brenner et al. \cite{Brenner_2002}. The approach takes into account the role of self-stress  and provides a quantitative estimate of the self-stress contribution to the overall bending and Gaussian stiffness. Indeed, the continuum model turns out to be the $\Gamma$-limit of the discrete graphene sheet, as proven in \cite{davini2018}. 

To understand the bending behavior of graphene is of the essence for several technological applications. The bending behavior controls the ripple formation and the performance of graphene nano-electro-mechanical devices \cite{huang2006,Lu_2009,Zhang_2011,Lindahl_2012,Tapaszto_2012,Kim_2012,Shi_2012,Hartmann_2013,Hajgato_2012,Wei_2013,Pacheco_2014,Favata_2016,Favata_2016b,Zelisko_2017,Genoese_2019}, and it is regarded as crucial in order to produce efficient hydrogen-storage devices \cite{Tozzini_2011,Goler_2013,Tozzini_2013}; moreover it can be instrumental to get inspiration for designing new metamaterials \cite{davini2017_2}.  Indeed, the intrinsic ripples are believed to be essential for the structural stability of the 2D graphene lattice and may have major impacts on the electronic and mechanical properties of graphene \cite{luHuang2009}.

In a recent review on \textit{Materials Today},  Deng and  Berry \cite{Deng_2016} give an overview of the hot problem of wrinkling, rippling, and crumpling, highlighting both formation mechanism and applications. The formation of these corrugations may have various explanations, see \cite{fasolino2007, nelsonPeliti1987, meyer2007}. Basically, the out-of-plane deformations (wrinkles and ripples) can significantly reduce
the magnitude of in-plane stresses generated, for instance, by defects, \cite{seungNelson1988, wang2013, zhang2014}. Zhang et al. \cite{zhang2014} adopted a generalized F\"oppl--von K\'arm\'an equation for a flexible solid membrane to describe ripples near defects such as disclinations (heptagons or pentagons) and dislocations (heptagon-pentagon dipoles) on graphene, and predicted the large scale graphene configurations under specific defect distributions. The paper closely follows a study of Seung and Nelson \cite{seungNelson1988}. Comparison with atomistic simulations indicates that the proposed model is capable to predict the atomic scale wrinkles near disclination/dislocation cores. The analysis shows that considering the buckling into 3-dimensional deformations is energetically more favorable than restricting to the in-plane ones. Similar defect-guided ripples in graphene were also simulated and discussed in the work of Wang et al. \cite{wang2013}.

With an eye toward wrinkling and ripple formation, here we deduce a continuum model of graphene for the case of finite out-of-plane displacements and  small in-plane deformations. We consider an array of C-atoms sitting at the nodes of a hexagonal lattice, and assume that the lattice interactions are governed by the Brenner's REBO potential of 2nd generation and that self-stress is present. Thus, the starting point is the same as in \cite{davini2017}, but the changes of edge lengths, wedge angles and dihedral angles are calculated by keeping  the quadratic term in the out-of-plane displacements, according to the form of the in-plane Green-Lagrange strain used in F\"oppl--von K\'arm\'an plate theory.

The computation of the approximated measures of strain is done in Section~3. Unlike Zhang et al., \cite{zhang2014}, that assume a triangular lattice for the continuum analysis as  done by Seung and Nelson in \cite{seungNelson1988}, here we use the real geometry of a hexagonal lattice. With due modifications, we adopt a harmonic approximation of the interatomic potential, which yields a splitting of the energy into {\it membrane} and {\it bending} parts, see Section~4. It follows that the bending part keeps the form already discussed in \cite{davini2017}, while the membrane part turns out to be affected by the non-linearity of the assumed in-plane Green-Lagrange strain. The continuum limit of the membrane energy is computed in Section~5 according to the formal approach followed in \cite{davini2017}. The founding assumption is that, to within a 
remainder tending to zero with the lattice size, the nodal displacements can be described by an average (macroscopic) displacement  and a relative {\it  shift displacement}  of the two Bravais lattices that give rise to the hexagonal periodicity. On minimizing the energy with respect to the shift variable, we formally determine a continuum model  of F\"{o}ppl--von K\'{a}rm\'{a}n type, whose constitutive coefficients are given in terms of the atomistic interactions.
%
%
%This conjecture anticipates the validity of a Cauchy--Born rule for the system plus the existence of a descriptor of the local adjustment of the two lattices, as suggested and justified within the framework of $\Gamma$-convergence theory by the analysis of \cite{davini2014}. 
%
A full validation of the obtained continuum limit within the scheme of $\Gamma$-convergence is left for future work.

\section{Kinematics and energetics } \label{sec:KIN}

\begin{figure}
	\centering
	\includegraphics[width=0.5\linewidth]{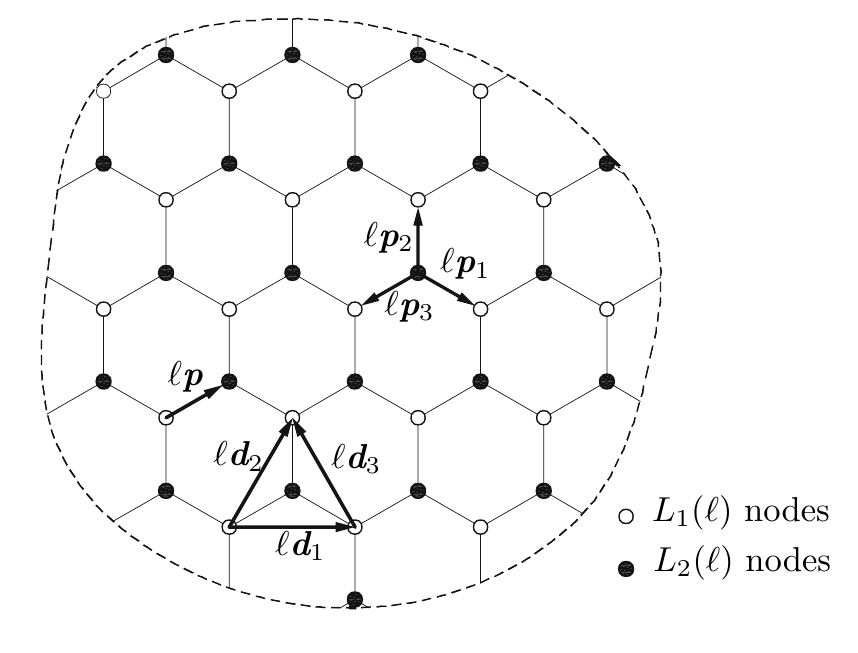}
	\caption{The hexagonal lattice}
	\label{fg:LATTICE}
\end{figure}

As reference configuration we use the $2$--{\it lattice} generated by two simple Bravais lattices 
\begin{equation}\label{eq:KIN_1}
\begin{array}{l} L_1(\ell) = \{ \mathbf{x} \in \mathbb{R}^2:
\mathbf{x} = n^1\ell\db_{1} + n^2\ell \db_{2} \quad \mbox{with} \quad (n^1,
n^2) \in \mathbb{Z}^2 \}, \\ L_2(\ell) = \ell\mathbf{p} + L_1(\ell),
\end{array}
\end{equation}
simply shifted with respect to one another by $\ell\pb$, see
Fig.~\ref{fg:LATTICE}.

In \eqref{eq:KIN_1}, $\ell$ denotes
the lattice size (the reference {\it interatomic distance}), while
$\ell\db_{\alpha}$ and $\ell\mathbf{p}$  respectively are the {\em lattice
	vectors} and the {\em shift vector}, with

\begin{equation}\label{eq:KIN_2}
\db_{1} =\sqrt{3}\eb_1, \quad \db_{2} = \frac{\sqrt{3}}{2}\eb_1+
\frac{3}{2}\eb_2 \quad \mbox{and} \quad \mathbf{p} =
\frac{\sqrt{3}}{2}\eb_1+ \frac{1}{2}\eb_2.
\end{equation}
The sides of the hexagonal cells in Figure~\ref{fg:LATTICE} stand for the bonds
between pairs of next nearest neighbor atoms and are represented by the
vectors
\begin{equation}\label{eq: KINENER 3}
\pb_\alpha = \db_\alpha - \pb \ \  (\alpha = 1, 2) \quad \mbox{and}
\quad \pb_3 = - \, \pb.
\end{equation}
For convenience we also set $$\db_3=\db_2-\db_1.$$

In what follows we denote by
\begin{equation}\label{eqnew}
\xb^\ell=n^1\ell\db_1+n^2\ell\db_2+m\ell\pb, \quad(n^1,n^2,m)\in\mathbb{Z}^2\times\{0,1\}
\end{equation}
the lattice points and label them by the triplets $(n^1,n^2,m)$: the points with $m=0$ belong to $L_1(\ell)$, while those in $L_2(\ell)$ correspond to $m=1$.

\begin{figure}
	\centering
	\includegraphics[width=0.9\linewidth]{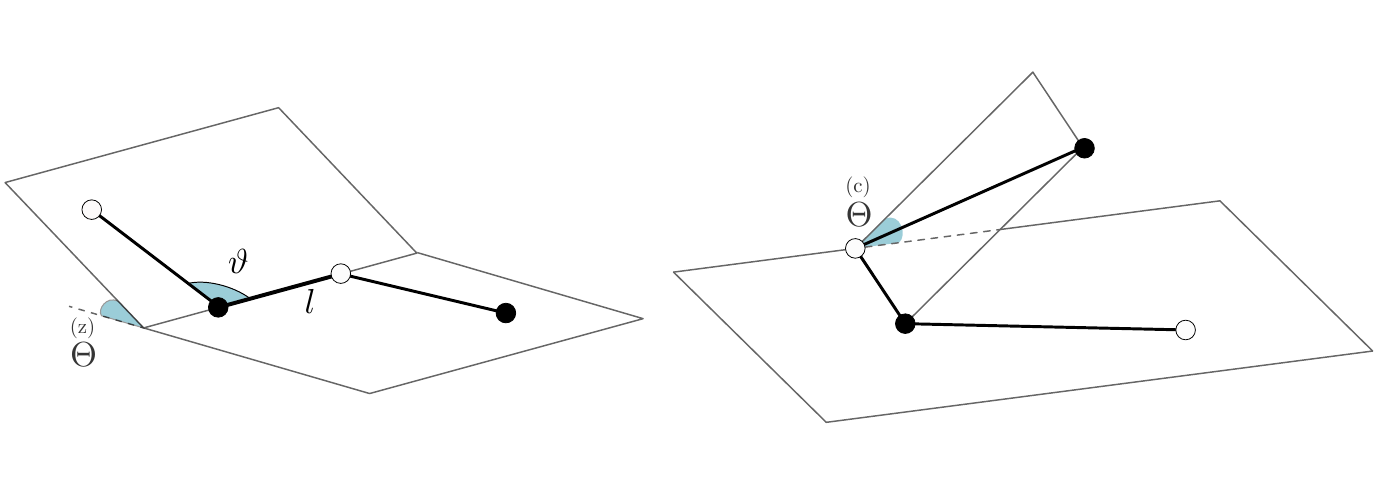}
	\caption{Edge bond $l$, wedge bond $\vartheta$,  Z-dihedron  $\Thz$ and a C-dihedron $\Thc$.}
	\label{fig:bonds}
\end{figure}

Graphene energetics depends on the description chosen to mimic atomic  interactions. Our model is based on the 2nd-generation Brenner potential \cite{Brenner_2002}, which is one of the most used in molecular dynamics simulations of graphene. Accordingly, the  binding energy $V$ of an atomic aggregate is given as a sum over nearest neighbors:
\begin{equation}\label{V}
V=\sum_i\sum_{j<i} V_{ij}\,, \quad V_{ij}=V_R(l_{ij})+b_{ij}(\vartheta_{hij},\Theta_{hijk})V_A(l_{ij}),
\end{equation}
where the individual effects of the \emph{repulsion} and \emph{attraction functions} $V_R(l_{ij})$ and $V_A(l_{ij})$, which model pair-wise interactions of the atoms $i$ and $j$ depending on their distance $l_{ij}$, are modulated by the \emph{bond-order function} $b_{ij}$; for a given bond chain  $h,i,j,k$ the function $b_{ij}$ depends in a complex manner on the angle between the edges $hi$ and $ij$ and on the dihedral angle between the planes spanned by $(hi,ij)$ and $(ij,jk)$. This potential reveals that,  in order to properly account  for the mechanical behavior of  graphene, it is necessary to consider three types of energetic contributions: 
\begin{enumerate}
	\item binary interactions between next nearest atoms  ({\it edge bonds}),
	\item three-body interactions between consecutive pairs of next nearest atoms  ({\it wedge bonds}),
	\item  and four-body interactions between three consecutive pairs of next nearest atoms ({\it dihedral bonds}).  There are two types of relevant dihedral bonds: the Z-dihedra, in which the edges connecting the four atoms form a Z-shape, and the C-dihedra, in which the edges form a C-shape (see Fig. \ref{fig:bonds}).
\end{enumerate}
Moreover, it is possible to show  \cite{Favata_2016} that  the angle  at ease between consecutive edges is greater than $\frac{2}{3}\pi$: this means that in the flat reference configuration  the graphene sheet is not  stress-free, and we will proper account for this feature.

\section{Approximated strain measures}\label{sec:STRAINS}
In this section we calculate the strain measures associated to a change of configuration described by a displacement field $\ub^\ell : (L_1(\ell)\cup L_2(\ell))\cap \Omega \to\mathbb{R}^3$. Since we have in mind to deduce a model with the same non-linearities as the F\"{o}ppl--von K\'{a}rm\'{a}n one, we write the displacements of the nodes in the form
\begin{equation}\label{uxi}
\ub^\ell(\xb^\ell)= \xi\vfb(\xb^\ell)+\xi^{1/2}\wf(\xb^\ell)\eb_3,
\end{equation}
where $\xi$ is a positive scalar measuring smallness, $\vfb:=\xi^{-1}\vb^\ell := \xi^{-1}(\eb_1\otimes\eb_1+\eb_2\otimes\eb_2)\ub^\ell$ and $\wf:= \xi^{-1/2}:=w^\ell \xi^{-1/2}\ub^\ell\cdot\eb_3$ stand for the in-plane and out-of-plane normalized displacements, respectively.

\subsection{Change of the edge lengths}
With $\delta l_i(\xb^\ell)$ we denote the change in length of the edge parallel to $\pb_i$ and starting from the lattice point $\xb^\ell\in (L_1(\ell)\cup L_2(\ell))\cap \Omega$. We fix our attention to lattice points in $L_2(\ell)$. Thus,
\begin{align*}
\delta l_i(\xb^\ell)&=|(\xb^\ell + \ell\pb_i + \ub^\ell(\xb^\ell+\ell\pb_i)) - (\xb^\ell +\ub^\ell(\xb^\ell))|-\ell\\
&=\ell|\pb_i + \frac{\ub^\ell(\xb^\ell+\ell\pb_i)- \ub^\ell(\xb^\ell)}{\ell} |-\ell.
\end{align*}
On introducing the notation
\begin{equation}
\Dl{i}f(\xb):=\frac{f(\xb+\ell\pb_i)-f(\xb)}{\ell},
\end{equation}
the axial strain measure can be recast as
\begin{equation}\label{nldl}
\frac{\delta\ell_i}{\ell_i}=|\pb_i+\Dl{i}\ub^\ell(\xb^\ell)|-1=|\pb_i+\xi\Dl{i}\vfb(\xb^\ell)+\xi^{1/2}\Dl{i}\wf(\xb^\ell)\eb_3|-1,
\end{equation}
where we have made use of \eqref{uxi}. The   expansion up to the first order in $\xi$ of the non-linear strain measure \eqref{nldl} is
\begin{equation}
\frac{\delta\ell_i}{\ell_i}=\xi \left(\Dl{i}\vfb(\xb^\ell)\cdot\pb_i+\frac12 |\Dl{i}\wf(\xb^\ell)|^2  \right)+o(\xi),
\end{equation}
which allows to define the  \textit{edge strain measure}, once rescaled-back by $\xi$:
\begin{equation}\label{epsi}
\varepsilon_i(\xb^\ell):=\Dl{i} \vb^\ell(\xb^\ell)\cdot\pb_i+\frac12 |\Dl{i}w^\ell(\xb^\ell)|^2 \qquad i=1,2,3,
\end{equation}
and $\delta\ell_i=\varepsilon_i\ell_i$.

\subsection{Change of the wedge angles}\label{sec:WANGLE}

\begin{figure}[H]
	\centering
	\includegraphics[scale=1]{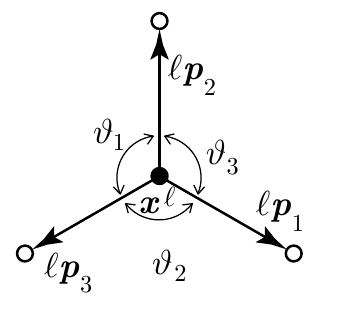}
	\caption{The wedge angles $\varthetat_i$.}
	\label{fig:angles_app}
\end{figure}

For each fixed node $\xb^\ell\in L_2(\ell))\cap \Omega$ we denote by $\varthetat_i(\xb^\ell)$ the angle of the wedge delimited by the edges $\pb_{i+1}$ and $\pb_{i+2}$; that is, the wedge angle opposite to the $i$-th edge (see Fig. \ref{fig:angles_app}). Here, $i, i+1$, and $i+2$ take values in $\{1,2,3\}$ and the sums should be interpreted mod 3: for instance, if $i=2$ then $i+1=3$ and $i+2=1$.
To keep the notation compact, we set
$$
\ub^\ell_{i}:=\ub^\ell(\xb^\ell+\ell\pb_{i}),\quad\mbox{and}\quad \ub^\ell_0:=\ub^\ell(\xb^\ell).
$$
Let 
\begin{equation}\label{q1}
\qb_{i+1}=\ell \left(\pb_{i+1}+\xi\Dl{i+1}\vfb(\xb^\ell)+\xi^{1/2}\Dl{i+1}\wf (\xb^\ell)\eb_3     \right )
\end{equation}
and 
\begin{equation}\label{q2}
\qb_{i+2}=\ell \left(\pb_{i+2}+\xi\Dl{i+2}\vfb(\xb^\ell)+\xi^{1/2}\Dl{i+2}\wf (\xb^\ell)\eb_3     \right )
\end{equation}
be the images of the edges parallel to $\pb_{i+1}$ and $\pb_{i+2}$ and starting at $\xb^\ell$. Then, the angle $\varthetat_i(\xi)=$ is given by
\begin{equation}\label{eq:wangle}
\varthetat_i(\xi)=\arccos\left(\frac{{\qb_{i+1}}\cdot{\qb_{i+2}}}{|{\qb_{i+1}}||{\qb_{i+2}}|}\right).
\end{equation} 

Now, in the light of \eqref{q1} and \eqref{q2}, we get
\begin{equation}
\frac{1}{\ell^2}\qb_{i+1}\cdot\qb_{i+2}=\pb_{i+1}\cdot\pb_{i+2}+\xi\Big(\pb_{i+1}\cdot\Dl{i+2}\vfb(\xb^\ell)+\pb_{i+2}\cdot\Dl{i+1}\vfb(\xb^\ell)+ \Dl{i+1}\wf(\xb^\ell)\Dl{i+2}\wf(\xb^\ell)     \Big ) +o(\xi),
\end{equation}
and
\begin{equation}
\begin{aligned}
&\frac{1}{\ell^2}|\qb_{i+1}|^2=1+\xi \big(|\Dl{i+1}\wf(\xb^\ell)|^2 +2\pb_{i+1}\cdot\Dl{i+1}\vfb(\xb^\ell)  \big)+o(\xi),\\
&\frac{1}{\ell^2}|\qb_{i+2}|^2= 1+\xi \big(|\Dl{i+2}\wf(\xb^\ell)|^2 +2\pb_{i+2}\cdot\Dl{i+2}\vfb(\xb^\ell)  \big) +o(\xi),\\
\end{aligned}
\end{equation}
so that the expansion up to the first order in $\xi$ of $\varthetat_i(\xi)$ yields
\begin{equation}
\varthetat_i(\xi)=\arccos\left(\frac{\pb_{i+1}\cdot\pb_{i+2}+a\xi+o(\xi)}{\sqrt{1+b\xi+o(\xi)}\,\sqrt{1+c\xi+o(\xi)}}\right)= \frac{2}{3}\pi - \frac{1}{2\sqrt{3}} (4a+b+c)\,\xi+o(\xi)  ,
\end{equation}
with
\begin{equation}
\begin{aligned}
&a:=\pb_{i+1}\cdot\Dl{i+2}\vfb(\xb^\ell)+\pb_{i+2}\cdot\Dl{i+1}\vfb(\xb^\ell)+ \Dl{i+1}\wf(\xb^\ell)\Dl{i+2}\wf(\xb^\ell),\\
&b:=|\Dl{i+1}\wf(\xb^\ell)|^2 +2\pb_{i+1}\cdot\Dl{i+1}\vfb(\xb^\ell),\\
&c:=|\Dl{i+2}\wf(\xb^\ell)|^2 +2\pb_{i+2}\cdot\Dl{i+2}\vfb(\xb^\ell).  
\end{aligned}
\end{equation}
This allows to define the \textit{wedge strain measure}, once the displacement is  rescaled-back by $\xi$:
\begin{multline}\label{psi2}
\psit_i(\xb^\ell)=\varthetat_i-\frac{2}{3}\pi=\\
-\frac{1}{\sqrt{3}}\Bigg( \Dl{i+1}\vb^\ell(\xb^\ell)\cdot\pb_{i+1} +\Dl{i+2}\vb^\ell(\xb^\ell)\cdot\pb_{i+2}  +2( \Dl{i+1}\vb^\ell(\xb^\ell)\cdot\pb_{i+2} +\Dl{i+2}\vb^\ell(\xb^\ell)\cdot\pb_{i+1}  )   +\\
\frac{1	}{2}\left( |\Dl{i+1}w^\ell(\xb^\ell)|^2+ |\Dl{i+2}w^\ell(\xb^\ell)|^2\right)     +2\Dl{i+1}w^\ell(\xb^\ell)\,\Dl{i+2}w^\ell(\xb^\ell)\Bigg)\,,
\end{multline}
for  $\xb^\ell\in L_2(\ell))\cap \Omega$. 
In particular, algebraic manipulations allow to conclude that
\begin{equation}\label{psisum}
\sum_{i=1}^3\psit_i(\xb^\ell)=-3\sqrt{3}\Bigg(\frac 13 \sum_{i=1}^3\Dl{i}w^\ell(\xb^\ell)\Bigg)^2.
\end{equation}

If we consider a lattice point belonging to $L_1(\ell)$, it is not difficult to see that
\begin{multline}\label{psi1}
\overset{(1)}{\psi}_i(\xb^\ell)=
\frac{1}{\sqrt{3}}\Bigg( \Dlo{i+1}\vb^\ell(\xb^\ell)\cdot\pb_{i+1} +\Dlo{i+2}\vb^\ell(\xb^\ell)\cdot\pb_{i+2}  +2( \Dlo{i+1}\vb^\ell(\xb^\ell)\cdot\pb_{i+2} +\Dlo{i+2}\vb^\ell(\xb^\ell)\cdot\pb_{i+1}  )   +\\
-\frac{1	}{2}\left( |\Dlo{i+1}w^\ell(\xb^\ell)|^2+ |\Dlo{i+2}w^\ell(\xb^\ell)|^2\right)     -2\Dlo{i+1}w^\ell(\xb^\ell)\,\Dlo{i+2}w^\ell(\xb^\ell)\Bigg)\,.
\end{multline}

\subsection{Change of the dihedral angles}
\begin{figure}[h]
	\centering
	\includegraphics[scale=1]{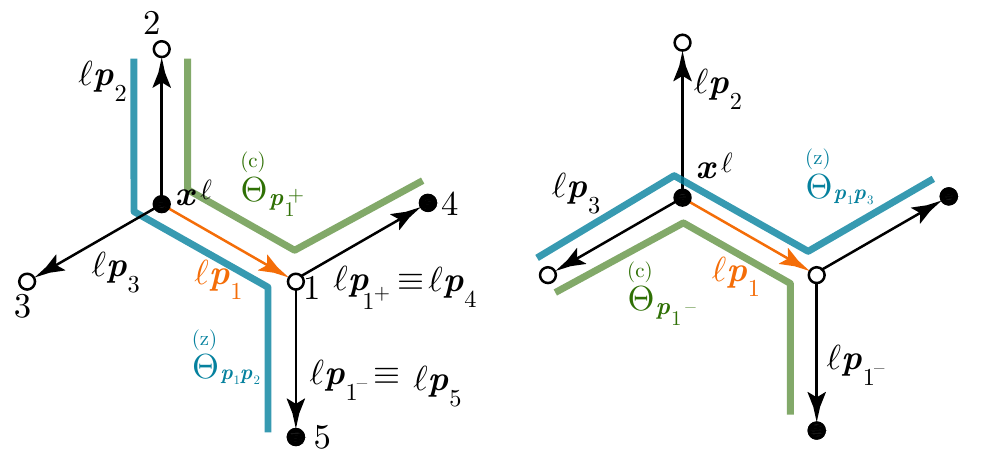}
	\caption{Left: C-dihedral angles $\Thc_{\pb_1^+}$ (green) and Z-dihedral angle $\Thz_{\pb_1\pb_2}$ (blue). Right: C-dihedral angles $\Thc_{\pb_1^-}$ (green) and Z-dihedral angle $\Thz_{\pb_1\pb_3}$ (blue).}
	\label{fig:cell_text}
\end{figure}

For each fixed node $\xb^\ell\in L_2(\ell)\cap \Omega$ and  for each edge parallel to $\pb_i$
and starting at $\xb^\ell$ we need to define four types of dihedral angles $\Thc_{\pb_i^+}(\xb^\ell), \Thc_{\pb_i^-}(\xb^\ell), \Thz_{\pb_i\pb_{i+1}}(\xb^\ell)$ and $\Thz_{\pb_i\pb_{i+2}}(\xb^\ell)$:

\begin{equation}
\begin{aligned}
&\cos\Thc_{\pb_i^+}=\frac{(\qb_i\times\qb_{i+1})\cdot(\qb_{i}\times\qb_{i^+})}{|\qb_i\times\qb_{i+1}||\qb_{i}\times\qb_{i^+}|},\\
&\cos\Thc_{\pb_i^-}=\frac{(\qb_{i+2}\times\qb_{i})\cdot(\qb_{i^-}\times\qb_{i})}{|\qb_{i+2}\times\qb_{i}||\qb_{i^-}\times\qb_{i})|},\\
&\cos\Thz_{\pb_i\pb_{i+1}}=\frac{(\qb_i\times\qb_{i+1})\cdot(\qb_{i^-}\times\qb_i)}{|\qb_i\times\qb_{i+1}||\qb_{i^-}\times\qb_i|},\\
& \cos\Thz_{\pb_i\pb_{i+2}}=\frac{(\qb_{i+2}\times\qb_i)\cdot(\qb_i\times\qb_{i^+})}{|\qb_{i+2}\times\qb_i||\qb_i\times\qb_{i^+}|},
\end{aligned}
\end{equation}
where	
\begin{equation}
\begin{aligned}
\qb_{i^+}=&\xb^\ell+\ell\pb_i-\ell\pb_{i+2}+\ub^\ell(\xb^\ell+\ell\pb_i-\ell\pb_{i+2})-\big( \xb^\ell+\ell\pb_i+\ub^\ell(\xb^\ell+\ell\pb_i) \big)\\
=&-\ell\pb_{i+2}+\ub^\ell_{i^+}-\ub^\ell_i, \qquad\qquad \ub^\ell_{i^+}:=\ub^\ell(\xb^\ell+\ell\pb_i-\ell\pb_{i+2}),\\
\qb_{i^-}=&\xb^\ell+\ell\pb_i-\ell\pb_{i+1}+\ub^\ell(\xb^\ell+\ell\pb_i-\ell\pb_{i+1})-\big( \xb^\ell+\ell\pb_i+\ub^\ell(\xb^\ell+\ell\pb_i) \big)\\
=&-\ell\pb_{i+1}+\ub^\ell_{i^-}-\ub^\ell_i, \qquad\qquad \ub^\ell_{i^-}:=\ub^\ell(\xb^\ell+\ell\pb_i-\ell\pb_{i+1})
\end{aligned}
\end{equation}
are the images of vectors $\ell\pb_{i^+}$ and $\ell\pb_{i^-}$ (see Fig. \ref{fig:cell_text}, for $i=1$), parallel to $\pb_{i+2}$ and $\pb_{i+1}$ and starting at the image of the point $\xb^\ell+\ell\pb_i$.

Also here, $i, i+1$, and $i+2$ take values in $\{1,2,3\}$ and the sums should be interpreted mod 3: for instance, if $i=3$ then $i+1=1$ and $i+2=2$.

The  C-dihedral angle $\Thc_{\pb_i^+}(\xb^\ell)$ is the angle corresponding to the C-dihedron with middle edge $\ell \pb_i$ and oriented as $\pb_i^\perp$,  while $\Thc_{\pb_i^-}(\xb^\ell)$ is the angle corresponding to the C-dihedron oriented opposite to $\pb_i^\perp$ (see Fig.~\ref{fig:cell_text} for $i=1$). 
The Z-dihedral angle $ \Thz_{\pb_i\pb_{i+1}}(\xb^\ell)$ 
corresponds to the Z-dihedron with middle edge $\ell \pb_i$ and the other two edges parallel to $\pb_{i+1}$
(see Fig.~\ref{fig:cell_text} for $i=1$).

To fix the ideas, we focus on the  dihedral angle $\Thz_{\pb_1\pb_2}$, sketched in Fig. \ref{fig:cell_text}; the other strains can be obtained in analogous manner.  

The first order approximation of the dihedral angle is all we need to evaluate the corresponding energy contribution. 

Let us introduce the vector 
$$
\qb_5=\ell\pb_5+(\vfb_5-\vfb_1)\xi+\wf_5\xi^{1/2},
$$
image of $\pb_5$ under the deformation. We have that 
\begin{equation}
\Thz_{\pb_1\pb_2}=\arccos\left(   \frac{\qb_1\times\qb_2}{|\qb_1\times\qb_2|}\cdot \frac{\qb_5\times\qb_1}{|\qb_5\times\qb_1|}  \right).
\end{equation}
Cumbersome computations yield:
\begin{equation}
\frac{\qb_1\times\qb_2}{|\qb_1\times\qb_2|}\cdot \frac{\qb_5\times\qb_1}{|\qb_5\times\qb_1|}  =1-\frac{2}{3\ell^2}(\wf_5-\wf_1+\wf_2-\wf_0)^2\xi+o(\xi).
\end{equation}
The expansion up to the first order in $\xi$ of $\Thz_{\pb_1\pb_2}$ yields
\begin{equation}
\Thz_{\pb_1\pb_2}(\xi)=\arccos \Big(1-a^2\xi+o(\xi)\big)=\sqrt{2 }a\, \xi^{1/2}+O(\xi^{3/2}), \quad a^2:=\frac{2}{3\ell^2}(\wf_5-\wf_1+\wf_2-\wf_0)^2.
\end{equation}
This allows to define the \textit{Z-dihedron strain measure}, once the displacement be  rescaled-back by $\xi$:
\begin{equation}
\Psiz_{\pb_1\pb_2}=\Thz_{\pb_1\pb_2}(\xi)-\arccos\left(   \frac{\pb_1\times\pb_2}{|\pb_1\times\pb_2|}\cdot \frac{\pb_5\times\pb_1}{|\pb_5\times\pb_1|}  \right)=\frac{2\sqrt{3}}{3\ell}\Big(w^\ell_5-w^\ell_1+w^\ell_2-w^\ell_0  \Big).
\end{equation}
For a generic Z-dihedral angle centered in $\pb_i$, we get
\begin{equation}
\begin{aligned}
& \Psiz_{\pb_i\pb_{i+1}}(\xb^\ell)=\frac{2\sqrt{3}}{3\ell}[w^\ell(\xb^\ell+\ell\pb_i-\ell \pb_{i+1})-w^\ell(\xb^\ell+\ell \pb_{i})+w^\ell(\xb^\ell+\ell \pb_{i+1})-w^\ell(\xb^\ell)],\\
& \Psiz_{\pb_i\pb_{i+2}}(\xb^\ell)=\frac{2\sqrt{3}}{3\ell}[w^\ell(\xb^\ell+\ell\pb_i-\ell \pb_{i+2})-w^\ell(\xb^\ell+\ell \pb_{i})+w^\ell(\xb^\ell+\ell \pb_{i+2})-w^\ell(\xb^\ell)].
\end{aligned}
\end{equation}
Analogous computations allows to determine the \textit{C-dihedron strain measure}
\begin{equation}
\begin{aligned}
&\Psic_{\pb_i^+}(\xb^\ell)=\frac{2\sqrt{3}}{3\ell}[2w^\ell(\xb^\ell)-w^\ell(\xb^\ell+\ell \pb_{i+1})+w^\ell(\xb^\ell+\ell\pb_i-\ell \pb_{i+2})-2w^\ell(\xb^\ell+\ell \pb_{i})],\\
&\Psic_{\pb_i^-}(\xb^\ell)=-\frac{2\sqrt{3}}{3\ell}[2w^\ell(\xb^\ell)-w^\ell(\xb^\ell+\ell \pb_{i+2})+w^\ell(\xb^\ell+\ell\pb_i-\ell \pb_{i+1})-2w^\ell(\xb^\ell+\ell\pb_i)].
\end{aligned}
\end{equation}

\section{Membrane and bending energy}\label{SEC:splitting}

The above calculations show that $\varepsilon$ and  $\psi$ depend upon both the in-plane and the out-of-plane  components of $\ub^\ell$, while $\Psic$, and $\Psiz$ depend upon the out-of-plane component of $\ub^\ell$; moreover, \eqref{psisum} shows that the sum of all $\psi$ depends on  the out-of-plane component of $\ub^\ell$. We introduce the following splitting of the energy into \textit{membrane} and \textit{bending} parts:
$$
\mathcal{U}_\ell=\mathcal{U}_\ell^{(m)}+\mathcal{U}_\ell^{(b)},   \quad  \mathcal{U}_\ell^{(m)}:=  \mathcal{U}_\ell^{(e)}+ \mathcal{U}_\ell^{(w)}\quad \mathcal{U}_\ell^{(b)}:=\Uc_\ell^{(s)} +\Uc_\ell^{(d)} 
$$   
defined by
\begin{equation}
\begin{aligned}
& \mathcal{U}_\ell^{(e)}:=\frac{1}{2} \, \sum_{\mathcal {E}} k^l (\delta\ell)^2=\frac{1}{2} \, \sum_{\mathcal {E}} k^l \ell^2\varepsilon^2 \\
& \mathcal{U}_\ell^{(w)}:=\frac{1}{2} \, \sum_{\mathcal {W}} k^\vartheta \ell^2 \, \psi^2 \\
&\Uc_\ell^{(s)} :=\tau_{0} \sum_{\mathcal {W}}\psi, \\ 
& \Uc_\ell^{(d)} :=\frac{1}{2} \, \sum_{\mathcal {Z}} k^\Zc \, \Psiz\phantomsection^2+\frac{1}{2} \, \sum_{\mathcal {C}} k^\Cc \, \Psic\phantomsection^2,
\end{aligned}
\end{equation}
where $\mathcal{U}_\ell^{(e)}$,  $\mathcal{U}_\ell^{(w)}$,  $\Uc_\ell^{(s)}$ and $ \Uc_\ell^{(d)}$ are the 
\textit{edge}, \textit{wedge}, \textit{self-}, and \textit{dihedral} energy, respectively.

The self-stress term is the outcome of the fact  that  the angle  at ease between consecutive edges is greater than $\frac{2}{3}\pi$: this means that in the flat reference configuration  the graphene sheet is not  stress-free, and $\tau_0$ represents the pre-stress couple (see \cite{Favata_2016}).  The constants $k^l$, $k^\vartheta$, $k^\Zc$, and $ k^\Cc $ can be deduced by means of the 2nd-generation Brenner potential. In the wedge energy we may interpret the presence of $\ell^2$ as a scaling of the constant 
$k^\vartheta$,  introduced to keep the energy finite as the lattice size $\ell$ goes to zero. We notice that the multiplication factor $\ell^2$ appears in all contributions to the membrane energy.

With the notation introduced in Section~\ref{sec:STRAINS} we now write the  energy more explicitly.
The edge energy can be written as 
\begin{equation}\label{Ueaa}
 \mathcal{U}_\ell^{(e)}=\frac{1}{2}\, k^{l} \ell^2 \sum_{\xb^\ell\in L_2(\ell)\cap\Omega}
 \sum_{i=1}^3   \varepsilon_i^2(\xb^\ell),
\end{equation}
while the wedge energy reads:
\begin{equation}\label{Uwaa}
 \mathcal{U}_\ell^{(w)}=\frac{1}{2}\, k^{\vartheta} \ell^2\sum_{\xb^\ell\in L_1(\ell)\cap\Omega}\sum_{i=1}^3  \overset{(1)}{\psi}_i\phantomsection{}^2(\xb^\ell)+ \frac{1}{2}\, k^{\vartheta}\ell^2 \sum_{\xb^\ell\in L_2(\ell)\cap\Omega}\sum_{i=1}^3  \overset{(2)}{\psi}_i\phantomsection{}^2(\xb^\ell).
\end{equation}
Analogously, the self-energy becomes
\begin{equation}\label{selfen1}
\mathcal{U}^{(s)}_\ell
= \sum_{\xb^\ell\in L_1(\ell)\cap\Omega}   \tau_0 \sum_{i = 
	1}^3\overset{(1)}{\psi}_i (\xb^\ell)+\sum_{\xb^\ell\in L_2(\ell)\cap\Omega}   \tau_0 \sum_{i = 
	1}^3\overset{(2)}{\psi}_i (\xb^\ell).
\end{equation}
We further split the dihedral energy  $\Uc_\ell^{(d)}$ in
$$
\Uc_\ell^{(d)}:=\mathcal{U}^{\mathcal{Z}}_\ell+\mathcal{U}^{\mathcal{C}}_\ell,
$$
where
\begin{equation}\label{enZ}
\mathcal{U}^{\mathcal{Z}}_\ell=\frac{1}{2}\, k^{\mathcal{Z}} \sum_{\xb^\ell\in L_2(\ell)\cap\Omega}
\sum_{i=1}^3 \Bigg(\Psiz_{\pb_i\pb_{i+2}}(\xb^\ell)\Bigg)^2+ \Bigg(\Psiz_{\pb_i\pb_{i+1}}(\xb^\ell)\Bigg)^2
\end{equation}
is the contribution of the Z-dihedra, and
\begin{equation}\label{enC}
\mathcal{U}^{\mathcal{C}}_\ell=\frac{1}{2} k^{\mathcal{C}} \, \sum_{\xb^\ell\in L_2(\ell)\cap\Omega}
\sum_{i=1}^3 \Bigg(\Psic_{\pb_i^+}(\xb^\ell)\Bigg)^2+ \Bigg(\Psic_{\pb_i^-}(\xb^\ell)\Bigg)^2
\end{equation}
is the contribution of the C-dihedra. 

We notice that the bending energy here obtained is the same as \cite{davini2017, davini2018}. In the following, we then focus on the membrane energy.

\section{The continuum limit of the membrane energy}\label{SEC:continuum}

In the previous sections the discrete  energy $\mathcal{U}_\ell$ was defined over the lattice $(L_1(\ell)\cup L_2(\ell))\cap\Omega$. By letting the lattice size $\ell$ go to zero the discrete set 
$(L_1(\ell)\cup L_2(\ell))\cap\Omega$ invades the domain $\Omega$, and the displacement functions
$w^\ell$ and $\vb^\ell$  will approach two functions $w$ and $\vb$ defined over $\Omega$. 
To  derive a continuous energy, defined over the domain $\Omega$, from
the discrete  energy $\mathcal{U}_\ell$
we need to specify the relations between $w^\ell$ and $w$ and between $\vb^\ell$ and $\vb$.

We assume  $w:\Omega \to \mathbb{R}$ to be twice continuously differentiable and to, almost, coincide with
$\wl$ over the lattice $(L_1(\ell)\cup L_2(\ell))\cap\Omega$. More precisely, we assume that
\begin{equation}\label{wCB}
\wl(\xb^\ell)=w(\xb^\ell)+o(\ell), \qquad \forall \xb^\ell\in (L_1(\ell)\cup L_2(\ell))\cap\Omega.
\end{equation}
If we think of $w$  and $\wl$ as the macroscopic and microscopic displacements, respectively, then \eqref{wCB} can be
thought as a Cauchy--Born rule. The assumption  \eqref{wCB} is motivated and essentially justified in \cite{davini2018}.

For the in-plane displacement we assume $\vb:\Omega \to \mathbb{R}^2$ to be continuously differentiable 
and 
\begin{equation}\label{vCB}
\begin{aligned}
\vbl(\xb^\ell)&=\vb(\xb^\ell)+o(\ell),   &\forall &\xb^\ell\in L_1(\ell)\cap\Omega,\\
\vbl(\xb^\ell)&=\vb(\xb^\ell)-\ell\sb(\xb^\ell)+o(\ell), \qquad &\forall &\xb^\ell\in L_2(\ell)\cap\Omega.
\end{aligned}
\end{equation}

Thus, over the lattice $L_1(\ell)$ we make the Cauchy--Born assumption for the in-plane displacement, while
over the lattice $L_2(\ell)$ this assumption is relaxed by introducing a ``shift displacement'' 
$\sb:\Omega \to \mathbb{R}^2$ that we assume to be continuously differentiable. Clearly, if the shift displacement is set equal to zero we have the Cauchy--Born rule over both lattices, but energetically it might be convenient to have a shift displacement $\sb$ different from zero. The minus sign in front of the term containing $\sb$ is introduced simply for later convenience. The assumption \eqref{vCB} is motivated and essentially justified in \cite{davini2014}. 

We note that for $\xb^\ell\in L_2(\ell)$ we have that $\xb^\ell+\ell\pb_i\in L_1(\ell)$, for $i=1,2,3$, and 
$$%\begin{eqnarray*}
\vbl(\xb^\ell+\ell\pb_i)=\vb(\xb^\ell+\ell\pb_i)+o(\ell)=\vb(\xb^\ell)+\ell\nabla\vb(\xb^\ell)\pb_i+o(\ell)=\vbl(\xb^\ell)+\ell\sb(\xb^\ell)+\ell\nabla\vb(\xb^\ell)\pb_i+o(\ell),
$$%\end{eqnarray*}
while for $\xb^\ell\in L_1(\ell)$ we have that $\xb^\ell-\ell\pb_i\in L_1(\ell)$, for $i=1,2,3$, and 
\begin{align}
\vbl(\xb^\ell-\ell\pb_i)&=\vb(\xb^\ell-\ell\pb_i)-\ell \sb(\xb^\ell-\ell\pb_i)+o(\ell)=\vb(\xb^\ell)-\ell\nabla\vb(\xb^\ell)\pb_i-\ell \sb(\xb^\ell)+o(\ell)\\ &=\vbl(\xb^\ell)-\ell\nabla\vb(\xb^\ell)\pb_i-\ell\sb(\xb^\ell)+o(\ell).
\end{align}
Thus,  
\begin{equation}\label{disv}
\begin{aligned}
\Dl{i}\vbl(\xb^\ell)&=\frac{\vbl(\xb^\ell+\ell\pb_i)-\vbl(\xb^\ell)}{\ell}=\nabla\vb(\xb^\ell)\pb_i+\sb(\xb^\ell)+o(1), &\xb^\ell\in L_2(\ell),\\
\Dlo{i}\vbl(\xb^\ell)&=\frac{\vbl(\xb^\ell-\ell\pb_i)-\vbl(\xb^\ell)}{\ell}=-\nabla\vb(\xb^\ell)\pb_i-\sb(\xb^\ell)+o(1), &
\xb^\ell\in L_1(\ell),
\end{aligned}
\end{equation}
and similarly for $\xb^\ell\in L_2(\ell)$ we have that
\begin{equation}\label{disw}
\Dl{i}\wl(\xb^\ell)=\frac{\wl(\xb^\ell+\ell\pb_i)-\wl(\xb^\ell)}{\ell}=\nabla w(\xb^\ell)\cdot\pb_i+o(1).
\end{equation}

For $i=1,2,3$ and for $\xb^\ell\in L_2(\ell)$ the edge strain measure defined in \eqref{epsi} writes as
\begin{align}
\varepsilon_i(\xb^\ell)&=\pb_{i}\cdot\Dl{i} \vbl(\xb^\ell)+\frac12 |\Dl{i}w(\xb^\ell)|^2 \\
&=\nabla\vb(\xb^\ell)\pb_i\cdot\pb_{i}+\sb(\xb^\ell)\cdot\pb_{i}+\frac12 |\nabla w(\xb^\ell)\cdot\pb_i|^2+o(1)\\
&=\El\vb(\xb^\ell)\pb_i\cdot\pb_{i}+\sb(\xb^\ell)\cdot\pb_{i}+\frac12 |\nabla w(\xb^\ell)\cdot\pb_i|^2+o(1)\\
&=[\El\vb(\xb^\ell)+\frac 12 \nabla w(\xb^\ell)\otimes \nabla w(\xb^\ell)]\cdot\pb_i\otimes\pb_{i}+\sb(\xb^\ell)\cdot\pb_{i}+o(1)\\
&=\Ev[\vb,w](\xb^\ell)\cdot\pb_i\otimes\pb_{i}+\sb(\xb^\ell)\cdot\pb_{i}+o(1)\label{epsi}
\end{align}
where
$$
\El\vb:=\frac{\nabla \vb +(\nabla \vb)^T}2, \qquad \Ev[\vb,w]:=\El\vb+\frac 12 \nabla w\otimes \nabla w,
$$
are the linearized and the von K\'arm\'an strain tensors, respectively.
Similarly, the wedge strain measure defined in \eqref{psi2} rewrites as
\begin{align}
\psit_i(\xb^\ell)&=\frac{-1}{\sqrt{3}}\Big(\big(\nabla \vb(\xb^\ell)\pb_{i+1}+\sb(\xb^\ell)\big)\cdot\pb_{i+1} +\big(\nabla \vb(\xb^\ell)\pb_{i+2}+\sb(\xb^\ell)\big)\cdot\pb_{i+2} \\
 & \hspace{1cm}+2\big(\nabla \vb(\xb^\ell)\pb_{i+1}+\sb(\xb^\ell)\big)\cdot\pb_{i+2} +2\big(\nabla \vb(\xb^\ell)\pb_{i+2}+\sb(\xb^\ell)\big)\cdot\pb_{i+1}\\
&\hspace{1cm}+\frac{1	}{2}\left( |\nabla w (\xb^\ell)\cdot \pb_{i+1}|^2+ |\nabla w (\xb^\ell)\cdot \pb_{i+2}|^2\right)     +2\nabla w (\xb^\ell)\cdot \pb_{i+1}\, \nabla w (\xb^\ell)\cdot \pb_{i+2}\Big)+o(1)\\
&=\frac{-1}{\sqrt{3}}\Big(\El \vb(\xb^\ell)\pb_{i+1}\cdot\pb_{i+1} +\El \vb(\xb^\ell)\pb_{i+2}\cdot\pb_{i+2}+ 4 \El \vb(\xb^\ell)\pb_{i+1}\cdot\pb_{i+2}\\
 & \hspace{1cm}+3\sb(\xb^\ell)\big)\cdot(\pb_{i+1} +\pb_{i+2})\\
&\hspace{1cm}+\frac{1}{2}\nabla w (\xb^\ell)\otimes \nabla w (\xb^\ell) \cdot \big( 
 \pb_{i+1}\otimes  \pb_{i+1}+ \pb_{i+2}\otimes  \pb_{i+2}+4 \pb_{i+1}\otimes  \pb_{i+2}\big)
\Big)+o(1)\\
 &=\frac{-1}{\sqrt{3}}\Big(\Ev[\vb,w](\xb^\ell)\cdot\Pb_i -3\sb(\xb^\ell)\cdot\pb_{i}\Big)+o(1),
\end{align}
where we set
\begin{equation}\label{Pgi}
\Pb_i:=\pb_{i+1}\otimes  \pb_{i+1}+ \pb_{i+2}\otimes  \pb_{i+2}+2 \pb_{i+1}\otimes  \pb_{i+2}+2 \pb_{i+2}\otimes  \pb_{i+1},
\end{equation}
and where we used the fact that $\pb_{i+1} +\pb_{i+2}=-\pb_i$.
A similar computation shows that, see \eqref{psi1},
\begin{equation}
\overset{(1)}{\psi}_i(\xb^\ell)=\frac{-1}{\sqrt{3}}\Big(\Ev[\vb,w](\xb^\ell)\cdot\Pb_i -3\sb(\xb^\ell)\cdot\pb_{i}\Big)+o(1).
\end{equation}

\begin{figure}
\centering
\begin{minipage}{.5\textwidth}
  \centering
  \includegraphics[width=.7\linewidth]{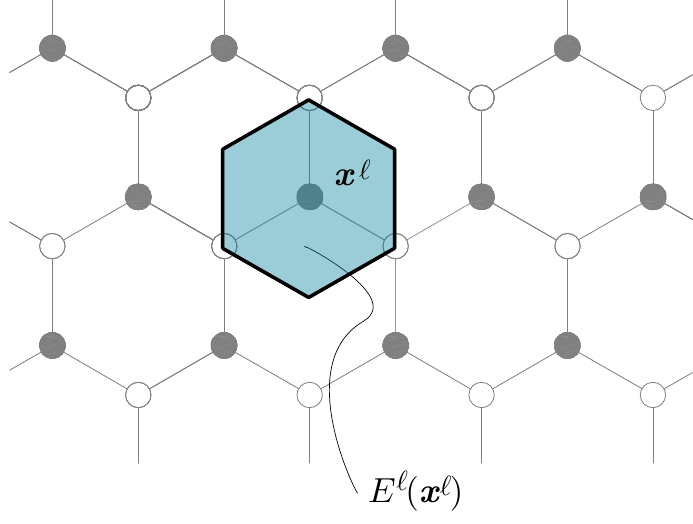}
  \caption{The hexagon $E^\ell(\xb^\ell)$.}
  \label{fighex}
\end{minipage}%
\begin{minipage}{.5\textwidth}
  \centering
  \includegraphics[width=.7\linewidth]{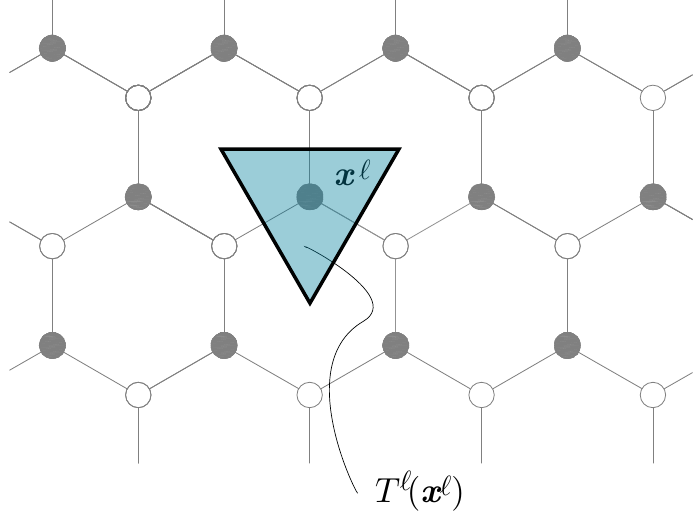}
  \caption{The triangle  $T^\ell(\xb^\ell)$}
  \label{figtri}
\end{minipage}
\end{figure}

We now compute the energies. The edge energy \eqref{Ueaa} becomes
\begin{align}
 \mathcal{U}_\ell^{(e)}&=\frac{1}{2}\, k^{l}\ell^2 \sum_{\xb^\ell\in L_2(\ell)\cap\Omega}
 \sum_{i=1}^3   \Big(\Ev[\vb,w](\xb^\ell)\cdot\pb_i\otimes\pb_{i}+\sb(\xb^\ell)\cdot\pb_{i}+o(1)\Big)^2\\
&=o(1)+\frac{1}{2}\, k^{l}\ell^2 \sum_{\xb^\ell\in L_2(\ell)\cap\Omega}
 \sum_{i=1}^3   \Big(\Ev[\vb,w](\xb^\ell)\cdot\pb_i\otimes\pb_{i}+\sb(\xb^\ell)\cdot\pb_{i}\Big)^2,
 \end{align}
 where the second equality is obtained by noticing that the number of points $\xb^\ell$ in $L_2(\ell)\cap\Omega$
is of order $1/\ell^2$.
 Let $|E^\ell(\xb^\ell)| = \ell^2 3\sqrt{3}/2$ be the area of the hexagon $E^\ell(\xb^\ell)$ of side $\ell$ centred at $\xb^\ell$, see Figure \ref{fighex}, and  let $\chi_{E^\ell(\xb^\ell)}(\xb)$ be the characteristic function of $E^\ell(\xb^\ell)$, i.e., the function equal to $1$ if $\xb \in E^\ell(\xb^\ell)$ and $0$ otherwise. The energy $ \mathcal{U}_\ell^{(e)}$ may be rewritten as
 \begin{align}
  \mathcal{U}_\ell^{(e)}&=o(1)+\frac{k^{l}}{3\sqrt{3}}\,  \sum_{i=1}^3 \int_{\Omega}\sum_{\xb^\ell\in L_2(\ell)\cap\Omega}
   \Big(\Ev[\vb,w](\xb^\ell)\cdot\pb_i\otimes\pb_{i}+\sb(\xb^\ell)\cdot\pb_{i}\Big)^2\chi_{E^\ell(\xb^\ell)}(\xb)\,d\xb \end{align}
and since the function $\sum_{\xb^\ell\in L_2(\ell)\cap\Omega}
   \big(\Ev[\vb,w](\xb^\ell)\cdot\pb_i\otimes\pb_{i}+\sb(\xb^\ell)\cdot\pb_{i}\big)^2\chi_{E^\ell(\xb^\ell)}(\xb)$ converges, as $\ell$ goes to zero, to $\big(\Ev[\vb,w](\xb)\cdot\pb_i\otimes\pb_{i}+\sb(\xb)\cdot\pb_{i}\big)^2$
   we have that
\begin{equation}
\lim_{\ell \to 0}  \mathcal{U}_\ell^{(e)}=\frac{k^{l}}{3\sqrt{3}}\,  \sum_{i=1}^3 \int_{\Omega}\big(\Ev[\vb,w](\xb)\cdot\pb_i\otimes\pb_{i}+\sb(\xb)\cdot\pb_{i}\big)^2\,d\xb=: \mathcal{U}_{0\sb}^{(e)}(\vb,w,\sb).
\end{equation}
 
By taking into account that the expression for  $\overset{(1)}{\psi}_i$ and
$ \overset{(2)}{\psi}_i$ are identical, the wedge energy \eqref{Uwaa} rewrites:

\begin{align}
 \mathcal{U}_\ell^{(w)}=o(1)+\frac{1}{2}\, k^{\vartheta}\ell^2 \sum_{\xb^\ell\in (L_1(\ell)\cup L_2(\ell))\cap\Omega}\sum_{i=1}^3  \frac{1}{{3}}\big(\Ev[\vb,w](\xb^\ell)\cdot\Pb_i -3\sb(\xb^\ell)\cdot\pb_{i}\big)^2
 \end{align}

%%%%%%%%%%

%%%%%%%%%%%
By introducing the triangles  $T^\ell(\xb^\ell)$ centered at $\xb^\ell$ of area $3\sqrt{3}\ell^2/4$ as depicted in
Figure \ref{figtri} and proceeding as above, we deduce that
\begin{equation}
\lim_{\ell \to 0}  \mathcal{U}_\ell^{(w)}=\frac{2k^{\vartheta}}{9\sqrt{3}}\,  \sum_{i=1}^3 \int_{\Omega}\big(\Ev[\vb,w](\xb)\cdot\Pb_i-3\sb(\xb)\cdot\pb_{i}\big)^2\,d\xb=:\mathcal{U}_{0\sb}^{(w)}(\vb,w,\sb).
\end{equation}
The limit of the membrane energy is
\begin{align}
\mathcal{U}_{0\sb}^{(m)}(\vb,w,\sb)&:=\lim_{\ell \to 0}  \mathcal{U}_\ell^{(m)}=\mathcal{U}_{0\sb}^{(e)}(\vb,w,\sb)+\mathcal{U}_{0\sb}^{(w)}(\vb,w,\sb)\\
&=\frac{k^{l}}{3\sqrt{3}}\,  \sum_{i=1}^3 \int_{\Omega}\big(\Ev[\vb,w](\xb)\cdot\pb_i\otimes\pb_{i}+\sb(\xb)\cdot\pb_{i}\big)^2\,d\xb\\
&\hspace{1cm}+\frac{2k^{\vartheta}}{9\sqrt{3}}\,  \sum_{i=1}^3 \int_{\Omega}\big(\Ev[\vb,w](\xb)\cdot\Pb_i-3\sb(\xb)\cdot\pb_{i}\big)^2\,d\xb\\
%%%%%%%%%
&= \int_{\Omega} \sum_{i=1}^3 \Big(\frac{k^{l}}{3\sqrt{3}}(\Ev[\vb,w](\xb)\cdot\pb_i\otimes\pb_{i})^2+
\frac{2k^{\vartheta}}{9\sqrt{3}}(\Ev[\vb,w](\xb)\cdot\Pb_i)^2\Big)
\\
&\hspace{14mm}+\sb\cdot   \sum_{i=1}^3 \Big(
\frac{2k^{l}}{3\sqrt{3}}\Ev[\vb,w](\xb)\cdot\pb_i\otimes\pb_{i}-\frac{4k^{\vartheta}}{3\sqrt{3}}\Ev[\vb,w](\xb)\cdot\Pb_i
\Big)\pb_{i}
\\
&\hspace{14mm}+  (\frac{k^{l}}{3\sqrt{3}}+\frac{2k^{\vartheta}}{\sqrt{3}}) \sum_{i=1}^3(\sb(\xb)\cdot\pb_{i}\big)^2\,d\xb\\
\end{align}
By using the relations
\begin{equation}\label{pi}
\pb_1=\frac{\sqrt 3}2\eb_1-\frac 12 \eb_2, \qquad \pb_2=\eb_2, \qquad \pb_3=-\frac{\sqrt 3}2\eb_1-\frac 12 \eb_2
\end{equation}
we find that
\begin{align}
\Ev\,\pb_1\cdot\pb_{1}&=\frac 34 (\Ev)_{11}+\frac 14 (\Ev)_{22}-\frac{\sqrt 3}2 (\Ev)_{12},\\
\Ev\,\pb_2\cdot\pb_{2}&=(\Ev)_{22},\\
\Ev\,\pb_3\cdot\pb_{3}&=\frac 34 (\Ev)_{11}+\frac 14 (\Ev)_{22}+\frac{\sqrt 3}2 (\Ev)_{12},
\end{align}
where $(\Ev)_{\alpha\beta}$ are the components of $\Ev=\Ev[\vb,w]$ with respect to the basis $\{\eb_i\}$.
A simple computation then shows that
\begin{align}\label{epp}
\sum_{i=1}^3 (\Ev\,\pb_i\cdot\pb_{i})^2&=\frac 18\Big(9(\Ev)_{11}^2+9(\Ev)_{22}^2+6(\Ev)_{11}(\Ev)_{22}+12(\El\vb)_{12}^2\Big),\\
&=\frac 38\Big(3(\tr \Ev)^2-4\det \Ev\Big).
\end{align}
Similarly we find 
\begin{gather}\label{sp}
\sum_{i=1}^3 (\Ev\cdot\Pb_i)^2=\frac{27}8\Big((\tr \Ev)^2-4\det \Ev\Big),\\
\sum_{i=1}^3 (\Ev\cdot\pb_i\otimes\pb_{i})\pb_i,=-\frac 32 (\Ev)_{12}\eb_1-\frac 34 \big((\Ev)_{11}-(\Ev)_{22}\big)\eb_2, \\
\sum_{i=1}^3 (\Ev\cdot\Pb_i)\pb_{i}=-\frac 92 (\Ev)_{12}\eb_1-\frac 94 \big((\Ev)_{11}-(\Ev)_{22}\big)\eb_2,\\
\sum_{i=1}^3 (\sb\cdot\pb_{i})^2=\frac 32 |\sb|^2.
\end{gather}

With these identities it follows that
\begin{align}
\mathcal{U}_{0\sb}^{(m)}(\vb,w,\sb)&= \int_{\Omega}  \frac{3(k^{l}+2k^{\vartheta})}{8\sqrt{3}}(\tr\Ev[\vb,w])^2
-\frac{k^{l}+6k^{\vartheta}}{2\sqrt{3}}\det\Ev[\vb,w]
\\
&\hspace{7mm}-\frac{k^{l}-6k^{\vartheta}}{2\sqrt{3}}\sb\cdot   \Big(
 2(\Ev[\vb,w])_{12}\eb_1+
\big(\Ev[\vb,w])_{11}-(\Ev[\vb,w])_{22}\big)\eb_2\Big)
\\
&\hspace{7mm}+  \frac{k^{l}+6k^{\vartheta}}{2\sqrt{3}}|\sb|^2\,d\xb.
\end{align}
The shift displacement $\sb$ enters into the membranal energy without any derivatives and therefore
it can be minimized once and for all. We readily see that the shift displacement that minimizes the energy is
\begin{equation}
\sb=\frac{k^{l}-6k^{\vartheta}}{2(k^{l}+6k^{\vartheta})} \Big(
 2(\Ev[\vb,w])_{12}\eb_1+
\big(\Ev[\vb,w])_{11}-(\Ev[\vb,w])_{22}\big)\eb_2\Big)
\end{equation}
and that
\begin{align}
\mathcal{U}_{0}^{(m)}(\vb,w)&:=\min_{\sb}\mathcal{U}_{0\sb}^{(m)}(\vb,w,\sb)
\\
&= \int_{\Omega}  \frac{3(k^{l}+2k^{\vartheta})}{8\sqrt{3}}(\tr\Ev[\vb,w])^2
-\frac{k^{l}+6k^{\vartheta}}{2\sqrt{3}}\det\Ev[\vb,w]
\\
&\hspace{7mm}-\frac{(k^{l}-6k^{\vartheta})^2}{8\sqrt{3}(k^{l}+6k^{\vartheta})} \big|
 2(\Ev[\vb,w])_{12}\eb_1+
\big(\Ev[\vb,w])_{11}-(\Ev[\vb,w])_{22}\big)\eb_2\big|^2\,d\xb.
\end{align}
By expanding the squares and reorganizing the terms we find:
\begin{align}
\mathcal{U}_{0}^{(m)}(\vb,w)&=\int_{\Omega}  \frac{3(k^{l}+6k^{\vartheta})}{8\sqrt{3}}(\tr\Ev[\vb,w])^2
-\frac{k^{l}+6k^{\vartheta}}{2\sqrt{3}}\det\Ev[\vb,w]
\\
&\hspace{7mm}-\frac{(k^{l}-6k^{\vartheta})^2}{8\sqrt{3}(k^{l}+6k^{\vartheta})} \Big((\tr\Ev[\vb,w])^2-4\det\Ev[\vb,w]\Big)\,d\xb,
\\
&= \int_{\Omega}  \frac{k^l(k^{l}+18k^{\vartheta})}{4\sqrt{3}(k^{l}+6k^{\vartheta})}(\tr\Ev[\vb,w])^2
-\frac{4\sqrt{3}k^{l}k^{\vartheta}}{k^{l}+6k^{\vartheta}}\det\Ev[\vb,w]
\,d\xb,
\end{align}
which is clearly an isotropic energy.
By means of the relation $\det \Ab=\big((\tr \Ab)^2-|\Ab|^2)/2$, which holds for every two by two matrix $\Ab$, we may write
\begin{align}
\mathcal{U}_{0}^{(m)}(\vb,w)&=
 \int_{\Omega} \frac 12 \lambda(\tr\Ev[\vb,w])^2
+\mu|\Ev[\vb,w]|^2
\,d\xb,
\end{align}
with

$$
\lambda:=\frac{k^l(k^{l}-6k^{\vartheta})}{2\sqrt{3}(k^{l}+6k^{\vartheta})}, \qquad \mu:=\frac{2\sqrt{3}k^{l}k^{\vartheta}}{k^{l}+6k^{\vartheta}}.
$$

\section{Conclusions}
We have proposed a non-linear continuum model for the mechanical behavior of graphene inferred from Molecular Dynamics potentials. Starting from a harmonic approximation of the energy as depicted by the 2nd-generation Brenner potential, we have found a discrete energy, depending on the displacement of each atom, which is sensitive to change of (i) the distance between two atoms (edge energy), (ii) the angle spanned by three subsequent atoms (wedge energy), (iii) two types of dihedral angles generated by the plane spanned by four subsequent atoms (C- and Z-dihedral energy). Thus, up-to-third neighbors interactions have been considered. Moreover, we have taken into account the presence of the self-stress, as predicted by the 2nd-generation Brenner potential (self-energy).

We have introduced a different scaling for in-plane and out-of-plane components, and this feature has produced a  coupling in the  strain measures at the discrete level. In particular, while the dihedral and the self-stress energies depend just on  the out-of-plane components, the edge and the wedges energies depend on both. These two latter contributions determine the \textit{membrane energy}, while the former two are part of the \textit{bending energy}. 
With the scaling here adopted, the discrete bending energy  turns out to be the same as that already considered in \cite{davini2017,davini2017_1,davini2018}; for this reason, we focused on the membrane energy.

The deduced discrete energy  is defined over the two Bravais lattices  generating the graphene sheet. By letting the size of the lattices to zero, they invade a continuum domain $\Omega$ and the  discrete displacement functions approach two continuous functions $w$ and $\vb$, representing the  in-plane and the out-of-plane continuum displacements, defined over $\Omega$. To obtain the continuum energy, it has been necessary to specify the relation between the discrete displacements and the corresponding continuum functions. To this end, motivated by \cite{davini2014}, we have made the Cauchy--Born assumption for the in-plane displacement defined over one of the Bravais lattices, while we have relaxed this assumption for the second lattice, by introducing a shift displacement $\sb$. 

With these assumptions, we have found a continuum membrane energy depending on $w$, $\vb$ and $\sb$. Since  the shift displacement $\sb$ enters into the membranal energy without any derivatives, we have minimized it once and for all; this has lead to a membrane energy depending just on $w$ and  $\vb$. 

Thus, on considering the bending energy already deduced in \cite{davini2017,davini2017_1,davini2018} and the membrane energy here found, we can state that the total continuum energy of graphene reads:
\begin{equation}
\mathcal{U}_{0}(\vb,w)=\mathcal{U}_{0}^{(m)}(\vb,w)+\mathcal{U}_{0}^{(b)}(w),
\end{equation}
where the membrane energy is given by
\begin{align}
\mathcal{U}_{0}^{(m)}(\vb,w)&=
\int_{\Omega} \frac 12 \lambda(\tr\Ev[\vb,w])^2
+\mu|\Ev[\vb,w]|^2
\,d\xb,
\end{align}
and the bending energy is
\begin{align}\label{entot}
\mathcal{U}^{(b)}_0(w)
=\frac 12 \int_\Omega \Dc ( \Delta w)^2+\Dc_G\det \nabla^2 w\,d\xb.
\end{align}
We find that 
\begin{equation}
\lambda=\frac{k^l(k^{l}-6k^{\vartheta})}{2\sqrt{3}(k^{l}+6k^{\vartheta})}, \quad \mu=\frac{2\sqrt{3}k^{l}k^{\vartheta}}{k^{l}+6k^{\vartheta}}, \quad \Dc= \frac{5\sqrt{3}}3 k^{\mathcal{Z}}+\frac{2\sqrt{3}}{3} k^{\mathcal{C}}-\frac{\tau_0}2, \quad 
\Dc_G=- \frac{8}5\frac{5\sqrt{3}}3 k^{\mathcal{Z}}-4\frac{2\sqrt{3}}{3} k^{\mathcal{C}},
\end{equation}
where  $k^l$, $k^{\vartheta}$, $k^{\mathcal{C}}$, $k^{\mathcal{Z}}$ and $\tau_0$ are the  constants entering the edge, wedge, C-dihedral, Z-dihedral, and self-, atomistic energies.

Within the limits of this formal deduction, we have  found that graphene can be  modeled as a classical  F\"{o}ppl--von K\'{a}rm\'{a}n plate, where the constitutive constants depend on  the atomistic interactions, as described by the 2nd-generation Brenner potential.

\section*{Acknowledgments}
A.F. acknowledges support from  Sapienza University of Rome through the   projects RP116154C92AF8A4  
``Multiscale Mechanics of 2D Materials: Modeling and Applications' and  RM11715C7F61C3E8 ``Shape morphing. From advanced differential geometry to applications in engineering and architecture''.

R.P.\  acknowledges support from the Universit\`a di Pisa through the project PRA\_2018\_61 ``Modellazione multi-scala in ingegneria strutturale''.

\end{document}